\documentclass{rspublic}
\RequirePackage[dvips]{graphicx}
\begin{document}

\title[Deuterium enhancement in H$_{3}$$^{+}$ in prestellar cores]{Deuterium enhancement in H$_{3}$$^{+}$ in prestellar cores}

\author[Charlotte Vastel et al.]{Charlotte Vastel$^{1}$ \& T.G. Phillips$^{1}$\\
P. Caselli$^{2}$, C. Ceccarelli$^{3}$, L. Pagani$^{4}$}
\affiliation{$^{1}$California Institute of Technology, USA\\
 $^{2}$ Osservatorio Astrofisico di Arcetri, Italy, $^{3}$LAOG, France, $^{4}$LERMA, France}

\label{firstpage}

\maketitle

\begin{abstract}{Astrochemistry, Deuterium, Interstellar}
Deuterium enhancement of monodeuterated species has been recognized
for more than 30 years as a result of the chemical fractionation that results from 
the difference in zero point energies of deuterated and
hydrogenated molecules. The key reaction is the deuteron exchange in
the reaction between HD, the reservoir of deuterium in dark
interstellar clouds, and the H$_{3}$$^{+}$ molecular ion, leading to
the production of the H$_{2}$D$^{+}$ molecule, and the low temperature
in dark interstellar clouds favors this production. Furthermore, the
presence of multiply deuterated species have incited our group to
proceed further and consider the subsequent reaction of H$_{2}$D$^{+}$
with HD, leading to D$_{2}$H$^{+}$ (first detected by Vastel et
al. 2004), which can further react with HD to produce D$_{3}$$^{+}$.
In prestellar cores, where CO was found to be depleted (Bacmann et
al. 2003), this production should be increased, as CO would normally
destroy H$_{3}$$^{+}$. The first model including D$_{2}$H$^{+}$ and
D$_{3}$$^{+}$ (Roberts, Herbst \& Millar 2003) predicted that these
molecules should be as abundant as H$_{2}$D$^{+}$ (see contribution by
H. Roberts). The first detection of the D$_{2}$H$^{+}$ was made
possible by the recent laboratory measurement by Hirao \& Amano (2003) for
the frequency of the fundamental line of the para-D$_{2}$H$^{+}$ (see
contribution by T. Amano).  Here we present observations of
H$_{2}$D$^{+}$ and D$_{2}$H$^{+}$ towards a sample of dark clouds and
prestellar cores and show how the distribution of
ortho-H$_{2}$D$^{+}$ (1$_{1,0}$-1$_{1,1}$) can trace the deuterium
factory in prestellar cores.  We also present how future instrumentation will 
improve our knowledge concerning the deuterium enhancement
of H$_{3}$$^{+}$.
\end{abstract}

\section{Introduction}

\begin{figure}
\centering
\includegraphics[scale=0.45]{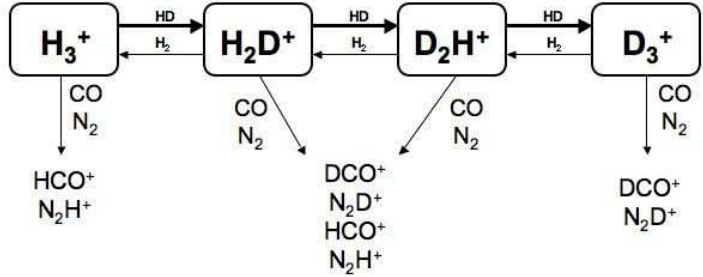}
\caption{Main reactions  involving the deuterated forms of the H$_{3}^{+}$ molecule. When CO and N$_{2}$ are 
depleted and the fractional ionization is $\leq$ 10$^{-7}$, the reactions with bold arrows are dominant.
\label{reactions}}
\end{figure}

Deuterium bearing species are good probes of the cold phases of
molecular clouds prior to star formation and many recent observations
point to the fact that their abundance relative to their hydrogenated
analogues can be larger, by a factor up to 10$^{5}$, than the solar
neighborhood value of $\sim$ 1.5 $\times$ 10$^{-5}$.  Therefore the
relative abundance of isotopologues does not measure the relative
abundances of the isotopes themselves.  The deuterium fractionation
has been evaluated in prestellar cores and low-mass protostars 
from observations of HCO$^+$ and
N$_2$H$^+$ (Butner et al. 1995; Williams et al. 1998; Caselli et
al. 2002; Crapsi et al. 2004, Crapsi et al. 2005), H$_2$CO (Loinard
et al. 2001; Bacmann et al. 2003), H$_2$S (Vastel et al. 2003), HNC
(Hirota et al. 2003), CH$_3$OH (Parise et al. 2004), and NH$_3$
(Roueff et al. 2000, Tin\'e et al. 2000).  The chemical fractionation
process in the gas--phase mainly arises from the difference between
the zero-point energies of H$_2$ and HD.  Almost incredibly, this can
lead to a detectable quantity of triply deuterated molecules like
ND$_3$ (Lis et al. 2002; van der Tak et al. 2002) and CD$_3$OH (Parise
et al. 2004).  Multiply deuterated methanol is thought to be formed
mainly on dust grain surfaces (Charnley et al. 1997) in regions where
the gas--phase [D]/[H] ratio is enhanced to values larger than
$\sim$0.1 (Parise et al. 2002). The high abundance found in the gas
phase for the D$_{2}$S also seem to favour the grain surface chemistry
scenario, when the [D/H] ratio is larger than 0.1 (Vastel et
al. 2003).  In molecular clouds, hydrogen and deuterium are
predominantly in the form of H$_2$ and HD respectively. So the
HD/H$_2$ ratio should closely equal the D/H ratio. Since the
zero-point energies of HD and H$_2$ differ by $\sim$ 410 K, the
chemical fractionation will favor the production of HD compared to
H$_2$.  In the dense, cold regions of the interstellar medium (T
$\sim$ 10 K), D will be initially nearly all absorbed into HD. The
abundant ion available for interaction is H$_3^+$, which gives
H$_2$D$^+$:
\begin{equation}
{\rm H_3^+ + HD  \longleftrightarrow H_2D^+ + H_2 + 230~K}
\end{equation}
The reverse reaction does not occur efficiently in the cold dense clouds where low--mass stars form, 
and where the kinetic temperature is always below 30~K, the ``critical'' temperature above which reaction 
(1.1) starts to proceed from right to left and limits the deuteration. Therefore, the degree of fractionation of 
H$_2$D$^+$ becomes non-negligible. This primary fractionation can then give rise to other 
fractionations and form D$_2$H$^+$ and D$_3^+$ as first suggested by Phillips \& Vastel (2003):
\begin{equation}
{\rm H_2D^+ + HD  \longleftrightarrow D_2H^+ + H_2 + 180~K}
\end{equation}
\begin{equation}
{\rm D_2H^+ + HD  \longleftrightarrow D_3^+ + H_2 + 230~K}
\end{equation}
We present in Figure \ref{reactions} the main reactions involving these molecules. Note that the effect 
of the dissociative recombination of H$_{3}$$^{+}$ is negligible because of the low electron density in 
such regions. Therefore, the reactions with CO or HD dominate the loss of H$_{3}$$^{+}$.

The dissociation of the deuterated forms of H$_3^+$ is then responsible of the enhancement in the 
[D]/[H] ratio. One specific parameter can enhance this process: the depletion of neutral species (in particular, 
the abundant CO) from the gas-phase (cf. Dalgarno \& Lepp 1984).
In fact, the removal of species which would normally destroy H$_3^+$ (e.g. CO; Roberts et al. 2000) means
that H$_3^+$ is more likely to react with HD and produce H$_2$D$^+$, D$_2$H$^+$ and D$_3^+$. The first model 
including D$_{2}$H$^{+}$ and D$_{3}^{+}$ (Roberts et al. 2003) predicted that these molecules should be as 
abundant as H$_{2}$D$^{+}$ (see also Flower et al. 2004). 

Gas phase species are expected to be depleted at the centers of cold,
dark clouds, since they tend to stick to the dust grains. A series of
recent observations has shown that, in some cases, the abundance of
molecules like CO decreases towards the core center of cold ($\le$ 10
K), dense ($\ge$ 2 $\times$10$^4$ cm$^{-3}$) clouds. L1544: Caselli et al. (1999); 
B68: Bergin et al. (2002); Oph D: Bacmann et al. (2003), Crapsi et al. (2005); 
L1521F: Crapsi et al. (2004); L183 (L134N): Pagani et al. (2005). These decreases in abundance have been
interpreted as resulting from the depletion of molecules onto dust
grains (see, e.g., Bergin et al. 1997, Charnley et al. 1997).  It
is now clear that these drops in abundance are typical of the majority
of dense cores. \\

\section{Observations}

For many years, H$_{2}$D$^{+}$ has been searched for (Phillips et
al. 1985; Pagani et al. 1992a; van Dishoeck et al. 1992; Boreiko \&
Betz 1993), and the advent of new submillimeter receivers led to the
detection of the 1$_{1,0}$--1$_{0,1}$ transition towards the young
stellar object, NGC 1333 IRAS 4A (Stark, van der Tak \& van Dishoeck
1999) although with relatively low signal strength, and in high
abundance towards the pre-stellar core L1544 (Caselli et al. 2003).  We
present in the following the recent advances on the deuterium
enhancement based on observations of pre-stellar cores performed at
CSO. The pre-stellar stage of star formation may be defined as the
phase in which a gravitationally bound core has formed in a molecular
cloud, and is evolving towards higher degrees of central condensation, but
no central protostellar object yet exists within the core.

\begin{figure}
\centering
\includegraphics[scale=0.45]{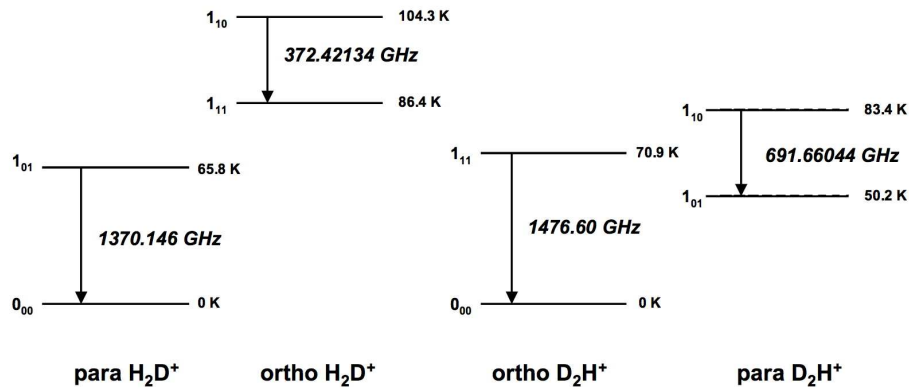}
\caption{Diagrams of the lowest energy levels for the H$_{2}$D$^{+}$ and D$_{2}$H$^{+}$ 
molecules.\label{transitions}}
\end{figure}

\subsection{16293E}

\begin{figure}
\centering
\includegraphics[scale=0.4]{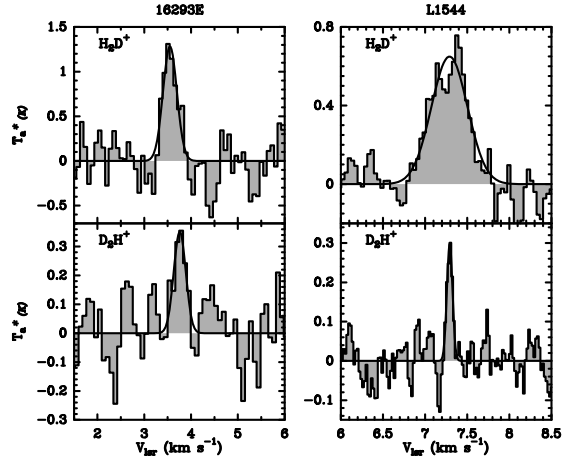}
\caption{Spectra of the ortho--H$_{2}$D$^{+}$ 1$_{1,0}$--1$_{1,1}$ and
para--D$_{2}$H$^{+}$ 1$_{1,0}$--1$_{0,1}$ transitions towards 16293E
(Vastel et al. 2004) and L1544 (Vastel et
al. \textit{submitted}).\label{16293e}}
\end{figure}

Encouraged by laboratory measurements (Hirao \& Amano 2003), Vastel et al. (2004) detected the para-D$_{2}$H$^{+}$ 
molecule in its ground transition at 692 GHz. By comparison with their H$_{2}$D$^{+}$ observations (see Figure 
\ref{16293e}), they found that in the prestellar core 16293E, the para-D$_{2}$H$^{+}$/ortho-H$_{2}$D$^{+}$ is $\sim$ 0.75. 
Both H$_{2}$D$^{+}$ and D$_{2}$H$^{+}$ molecules have ortho and para forms, corresponding to the spin states 
of the protons (for H$_{2}$D$^{+}$) or deuterons (for D$_{2}$H$^{+}$). In order to compare the modeled abundances 
with the observations of one spin state only, it is critical to know the ortho--to--para ratio for these two molecules. 
Under LTE conditions, at temperature T, the relative populations of the lowest ortho (1$_{1,1}$) and para (0$_{0,0}$) 
levels of H$_{2}$D$^{+}$ would be:\\
\begin{equation}
{\rm \frac{n(1_{1,1})}{n(0_{0,0})}=9 \times exp(-\frac{86.4}{T})}
\end{equation}
and the relative populations of the lowest ortho (0$_{0,0}$) and para (1$_{0,1}$) levels of D$_{2}$H$^{+}$ would be:\\
\begin{equation}
{\rm \frac{n(1_{0,1})}{n(0_{0,0})}=\frac{9}{6} \times exp(-\frac{50.2}{T})}
\end{equation}
It results that, at 8 K, the H$_{2}$D$^{+}$ ortho-to-para ratio would
be $\sim$ 1.8 10$^{-4}$ and the D$_{2}$H$^{+}$ para-to-ortho ratio
would be $\sim$ 2.8 10$^{-3}$. The ortho form of H$_{2}$D$^{+}$ is
produced mainly in reactions of the para form with ortho-H$_{2}$
(e.g. Gerlich, Herbst \& Roueff 2002).  Therefore, its high o/p ratio
is attributable to the relatively high ortho-H$_{2}$ abundance as first noticed by Pagani et al. 
(1992b). Because the o/p ratio is not thermalized at the low
temperature considered here, neither is the o/p H$_{2}$D$^{+}$ ratio.  
This can be illustrated in Flower, Pineau des
For\^ets, Walmsley (2004) model where, at temperatures lower than 10
K, a hydrogen density of 2 $\times$ 10$^{6}$ cm$^{-3}$ and a grain
size of 0.1 $\mu$m, the o/p-H$_{2}$D$^{+}$ reaches unity and the
p/o-D$_{2}$H$^{+}$ value is about 0.1. Consequently the D$_{2}$H$^{+}$
abundance should be about 4 times higher than the H$_{2}$D$^{+}$
abundance in the 16293E core.

This study supported chemical modelling and the inclusion of multiply deuterated species 
(Roberts et al. 2003; Walmsley et al. 2004; Roberts et al. 2004; Flower et al. 2005; Aikawa et al. 2005). 

\subsection{L1544}

\begin{figure}
\centering
\includegraphics[scale=0.4,angle=270]{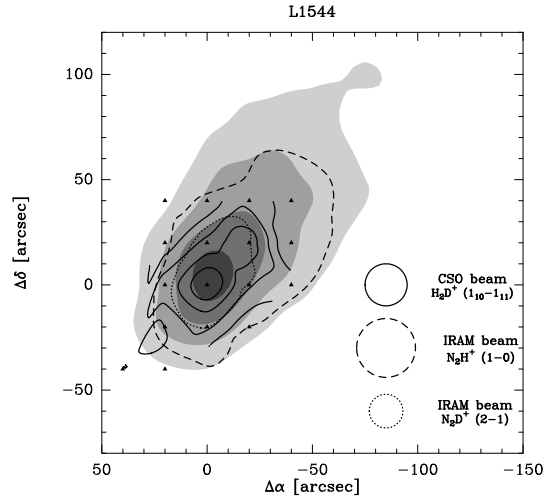}
\caption{Integrated intensity maps of H$_{2}$D$^{+}$ (1$_{1,0}$-1$_{1,1}$), N$_{2}$H$^{+}$ (1--0) and 
N$_{2}$D$^{+}$ (2-1) superposed on the 1.3 mm continuum emission map of the pre-stellar core L1544. 
Contour levels are 30\%, 50\%, 70\% and 90\% of the peak for H$_{2}$D$^{+}$, and 50\% of the peak for 
N$_{2}$H$^{+}$ and N$_{2}$D$^{+}$. The observed positions in H$_{2}$D$^{+}$ are reported as triangles. 
\label{L1544_map}}
\end{figure}

In one of the most heavily CO--depleted prestellar cores, L1544 where f$_{D}$ $\sim$ 10,  
Caselli et al. (2003) detected a strong (T$_{mb}$ $\sim$ 1~K) ortho-H$_2$D$^+$(1$_{01}$-1$_{11}$) line 
(see Figure \ref{16293e}), and concluded 
that H$_2$D$^+$ is one of the main molecular ions in the central region of this core. Vastel et al. (submitted) mapped 
the area around the dust peak position (see Figure \ref{L1544_map}) and found that the H$_{2}$D$^{+}$ distribution 
closely follow the dust continuum in contrast to the CO molecule that appears to be depleted. Also, the para-D$_{2}$H$^{+}$ 
1$_{1,0}$--1$_{0,1}$ transition  was observed at the dust peak position (see Figure \ref{16293e}). However, the 
linewidth is about 3 times lower than the expected thermal linewidth for a kinetic temperature of 7 K, as predicted by 
dust temperature measurement. In absence of a possible explanation for this profile, this observation has been considered 
as a tentative detection only.   

\begin{figure}
\centering
\includegraphics[scale=0.4,angle=0]{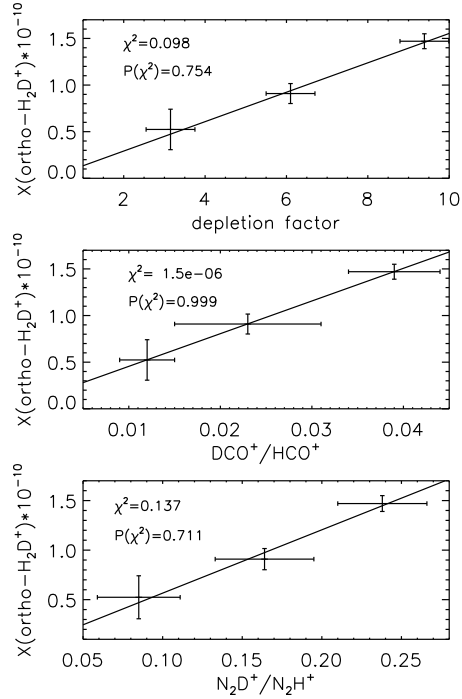}
\caption{ Variation of the observed ortho-H$_{2}$D$^{+}$ as a function of the depletion factor, DCO$^{+}$/HCO$^{+}$ ratio 
and N$_{2}$D$^{+}$/N$_{2}$H$^{+}$ ratio in the pre-stellar core L1544. The $\chi^{2}$ parameter and its probability are reported.
\label{variations}}
\end{figure}

 They show the correlation between the ortho-H$_{2}$D$^{+}$ abundances at the 0$^{\prime\prime}$, 
 $\pm$ 20$^{\prime\prime}$ and $\pm$ 40$^{\prime\prime}$ distance from the dust peak and the CO 
 depletion factor, the DCO$^{+}$/HCO$^{+}$ ratio and the N$_{2}$D$^{+}$/N$_{2}$H$^{+}$ ratio (see Figure 
 \ref{variations}). As intuitively expected, the ortho-H$_{2}$D$^{+}$ abundance appears to be well correlated with 
 the CO depletion. Also, the fractionation ratios for the N$_{2}$H$^{+}$ and HCO$^{+}$ molecules increase 
 linearly with the ortho-H$_{2}$D$^{+}$ abundance. The surprisingly high confidence level for the correlations between 
 H$_{2}$D$^{+}$ and the degree of deuteration in the HCO$^{+}$ and N$_{2}$H$^{+}$ molecules confirms that 
 H$_{2}$D$^{+}$ dominates the fractionation of these molecules at low temperatures. Therefore, in the pre-stellar core 
 L1544, D$_{2}$H$^{+}$ and D$_{3}$$^{+}$ should not intervene in these fractionations (see Figure \ref{reactions}).

\subsection{L183 (L134N)}

\begin{figure}
\centering
\includegraphics[scale=0.5]{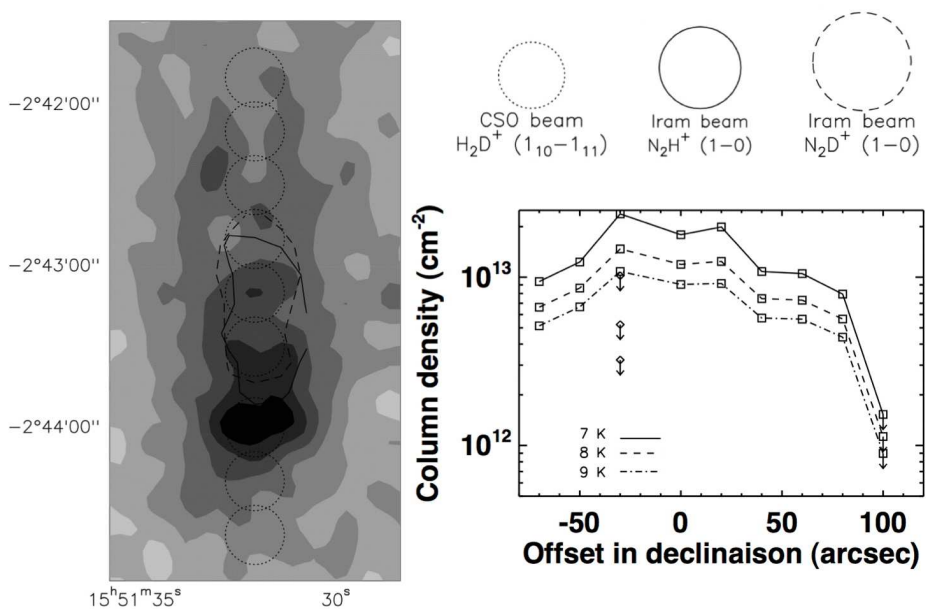}
\caption{Integrated intensity maps of N$_{2}$H$^{+}$ (1--0) and 
N$_{2}$D$^{+}$ (1-0) superposed on the dust emission map of the pre-stellar core L183. 
Contour levels  90\% of the peak for N$_{2}$H$^{+}$ and 85\% for N$_{2}$D$^{+}$. 
The o-H$_{2}$D$^{+}$ column densities (squares) and the upper limit on p-D$_{2}$H$^{+}$ (diamonds) are reported 
for T$_{ex}$~=~7, 8, 9 K. \label{l134n}}
\end{figure}

Another example is the cold, dark, starless cloud core L183, where we present observations of the 
ortho-H$_2$D$^+$ (1$_{10}$-1$_{11}$) ground transition (Vastel, Pagani et al. {\it in preparation}). 
We traced the central ridge and the central peak of L183 (defined in Pagani et al. 2003, 2004) 
together with detailed maps of several
transitions of N$_2$H$^+$ and N$_2$D$^+$ obtained at IRAM and at CSO (see Figure \ref{l134n}).
N$_2$H$^+$ and N$_2$D$^+$ do not trace the dust peak and thus are depleted in the
most inner part of the cloud (Pagani et al. 2005) while H$_2$D$^+$, as expected 
from theory and other source observations (as in the case of L1544) does peak at the dust peak. Surprisingly,
the H$_2$D$^+$ is very extended, spanning 150$^{\prime\prime}$ in declination and presents
a second intensity peak, of similar strength to the  main peak, 50$^{\prime\prime}$ 
north of it. This second peak has no local dust counterpart.

We also searched for para-D$_2$H$^+$ (1$_{10}$-1$_{01}$) at 692 GHz towards the main 
dust peak but have found no emission so far. The para-D$_2$H$^+$/ortho-H$_2$D$^+$ ratio is thus 
lower than 0.4, two times less than in the case of 16293E and at least 2 times less than in the case of 
L1544.

\section{Perspectives}

Recently Caselli et al. (\textit{in preparation}) performed a survey in H$_{2}$D$^{+}$ over a sample of 
prestellar cores, and some protostars. Three quarters of these sources present a strong emission (between 
0.5 to 1 K in main beam temperature). In dark clouds affected by molecular depletion, the deuterated 
forms of the molecular ion H$_3^+$ 
are unique tracers of the core nucleus, the future stellar cradle.  Thus, their study becomes fundamental to unveil the 
initial conditions of the process of star formation (kinematics, physical and chemical structure of pre--stellar cores). 
Table 1 lists some of the major telescopes and interferometers that can be used for 
the study of H$_{2}$D$^{+}$ chemistry in prestellar, proto-planetary disks and protostars. 
Furthermore, the access to the para-H$_{2}$D$^{+}$ and ortho-D$_{2}$H$^{+}$ transitions with new 
submillimeter receivers on ground based and space telescopes will enable to determine precisely 
the ortho--to--para ratios for both molecules.

\begin{table}
\caption{Current and future facilities for the chemistry of H$_{2}$D$^{+}$ and D$_{2}$H$^{+}$.}
\tiny
\begin{tabular}{ccccccc}
\hline
Name                 &   Aperture            & Available     &\multicolumn{2}{c}{H$_{2}$D$^{+}$}          & \multicolumn{2}{c}{D$_{2}$H$^{+}$}\\
          &            &  &1$_{1,0}$--1$_{1,1}$ &1$_{0,1}$--0$_{0,0}$ &1$_{1,0}$--1$_{0,1}$  & 1$_{1,1}$--0$_{0,0}$\\                          
          &            &  & (372.4 GHz) & (1.37 THz) & (691.7 GHz) & (1.48 THz)\\
\hline
CSO                  &  10.4 m                & Y                & Y & N & Y & N\\
JCMT                &  15 m                   & Y                & HARP B & N & Y & N \\
SOFIA               &   2.5 m                  &  2007           & N & Casimir & Casimir & Casimir \\
                          &                            &                    &     &  GREAT (CONDOR) &    &   GREAT (CONDOR)\\
Herschel  (HIFI)  &   3.5 m                  &  2007           & N & N & Y & Y\\
ALMA                & 50 $\times$ 12 m &  2010           & Y & N & Y & N\\
APEX                 &   12 m                  &  2005-2006   & Y & CONDOR & Y & CONDOR\\
\hline
\end{tabular}
\end{table}

\label{lastpage}
\end{document}